\documentclass{ws-procs9x6-cpt19}
\begin{document}

\newcommand{\refeq}[1]{(\ref{#1})}
\def\etal {{\it et al.}}

\title{Testing Lorentz violation by the comparison of atomic clocks}

\author{Xiao-yu Lu, Yu-Jie Tan, and Cheng-Gang Shao}

\address{ MOE Key Laboratory of Fundamental Physical Quantities Measurements ${\rm{\& }}$ Hubei Key Laboratory
of Gravitation and Quantum Physics, \\ PGMF and School of Physics, Huazhong University of Science
and Technology, \\ Wuhan 430074, People¡¯s Republic of China}

\begin{abstract}
A more complete theoretical model of testing Lorentz violation by the comparison of atomic clocks is developed in the
Robertson-Mansouri-Sexl kinematic framework. As this frame postulates the deviation of the coordinate transformation from the Lorentz transformation, from the viewpoint of the transformation violations on time and space, the frequency shift effect in the atomic clock comparison can be explained as two parts: time-delay effect $\alpha \frac {v^2}{c^2}$ and structure effect $-\frac {\beta+2\delta}{3} \frac {v^2}{c^2}$. Standard model extension is a widely used dynamic frame to characterize the Lorentz violation, in which a space-orientation dependence violating background field is added as the essential reason for the Lorentz violation effect. Compared with the RMS frame which only indicates the kinematic properties with the coordinate transformation, this dynamic frame provides a more complete and clear description for the possible Lorentz violation effect.
\end{abstract}

\bodymatter

\section{Introduction}

Lorentz invariance (LI) is the fundamental symmetry of spacetime, which postulates the experimental result independent on the motion state of the apparatus \cite{1}. As LI is at the foundation of both the Standard model of particles physics and general relativity, its related research is an important subject in the physics science. Here, we studied the LI effect in Robertson-Mansouri-Sexl (RMS) framework,\cite{4,5} and made a simple comparison with that in Standard-Model Extension (SME) framework.\cite{2,3} RMS framework considers the speed of light anisotropic, and also postulates there is a preferred universal frame in which light propagates conventionally as measured using a set of rods and clocks. For these RMS rods and clocks, they are isotropic and the photon is anisotropic, while for the SME rods and clocks, the case is opposite. Since LI violation results in the difference of transition frequency, the atomic clock comparison is a good means to test this violating effect, and we focus on analyzing this violation between comparisons.

\section{Lorentz violation of atomic clock comparison}

The violation of LI is described in RMS kinematic framework as the deformation of Lorentz transformation, and it postulates the existence of a preferred frame $\Sigma$, that is, cosmological microwave background (CMB) frame. If the laboratory reference frame $S$ has the velocity $v$ with respect to $\Sigma$ frame, the transformation between these two frames can be written as \cite{4,5}
\begin{eqnarray}\label{n1}
&&t=aT + \vec {\varepsilon } \cdot \vec {x},\;\;\;x=b(X-\emph{v}T),\;\;\;y=dY,\;\;\;z=dZ
\end{eqnarray}
with $a(v)=1 + (\alpha  - \frac{1}{2})\frac{{{v^2}}}{{{c^2}}}$, $b(v)=1 + (\beta  + \frac{1}{2})\frac{{{v^2}}}{{{c^2}}}$ and $d(v) = 1 + \delta \frac{{{v^2}}}{{{c^2}}}$,\cite{6} which returns to the Lorentz transformation with $\alpha=\beta=\delta=0$. For the comparison of clock frequencies, the violation of LI in the $S$ frame can be detected through measuring the anisotropic of light speed. Analyzing light-clock and atom-clock comparisons in Fig.\,{1(a)} and {1(b)},\cite{7, 8} the frequency shift signal of clock-comparison experiment contains the time-delay and structure effects \cite{9}
\begin{eqnarray}\label{n2}
\Delta_{LV}=\alpha\frac{v^2}{c^2}-\frac{\beta+2\delta}{3}\frac{v^2}{c^2},
\end{eqnarray}
where the first part means time delay and the other is structure effect.

\begin{figure}
\includegraphics[width=4in]{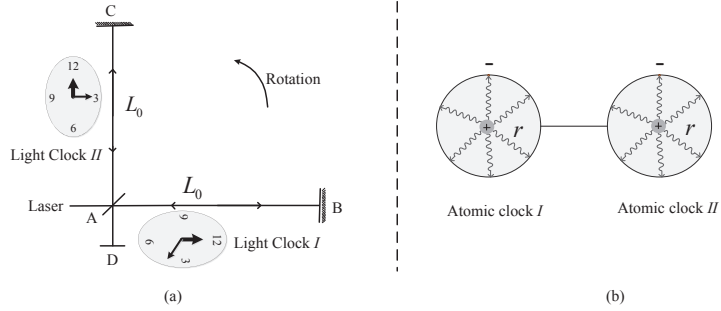}
\caption{Two kinds of clock-comparison experiments. (a) The comparison of light clocks: it is similar to a Michelson interferometer with each arm length $L_0$, where each interference arm can be considered as a light clock. (b) The comparison of atomic clocks: two atomic clocks are located in different places.}
\label{aba:fig1}
\end{figure}

The SME framework provides a general theoretical framework for studying the violation of LI, such as the violation in photon, matter, gravity sectors and so on. Compared with RMS framework, SME framework provides a vast parameter space. In this framework, a LI violating background field is postulated, and different coordinate systems are linked by the Lorentz transformation. Therefore, for the atomic clock comparison, there are no LI violation in photon sector, and the main violating effect is embodied in the matter sector.\cite{10} RMS formalism can be regarded as a special limit of SME LI violation formalism.

\section{conclusion}

Based on the theoretical analysis of testing LI violation by light-clock comparison in RMS frame, we studied the test by atomic clock comparison, and the result indicates LI violation effect includes the time-delay and structure effects. In addition, we also make a simple explanation for the different indications of LI violation for atomic clock comparison in the RMS and SME frames. For RMS framework, the violating effect arises from the deformation of Lorentz transformation, while for the SME frame, it results from the existed violating background field.

\section{Acknowledgments}

This work is supported by the Post-doctoral Science Foundation of China (Grant Nos. 2017M620308 and 2018T110750).

\end{document}